\documentclass[12pt,preprint]{aastex}

\usepackage[british]{babel}
\usepackage{url}

\usepackage{datetime}

\ddmmyyyydate

\shorttitle{Coronal Propagating Fronts by SDO/AIA} \shortauthors{Nitta
et al.}

\begin{document}



\title{Large-scale Coronal Propagating Fronts in Solar Eruptions as
  Observed by the Atmospheric Imaging Assembly on Board the Solar
  Dynamics Observatory\,--\,An Ensemble Study}


\author{Nariaki V. Nitta\altaffilmark{1}, Carolus
  J. Schrijver\altaffilmark{1},  Alan M. Title\altaffilmark{1} 
  and Wei Liu\altaffilmark{1,2}}

\altaffiltext{1}{Lockheed Martin Advanced Technology Center,
  Dept/A021S, B/252, 3251 Hanover Street, Palo Alto, CA 94304, USA}
\altaffiltext{2}{W. W. Hansen Experimental Physics Laboratory,
  Stanford University, Stanford, CA 94305, USA}




\begin{abstract}

This paper presents a study of a large sample of 
global disturbances in the solar corona with characteristic
propagating fronts as intensity enhancement, similar to the phenomena 
that have often been referred to as EIT waves 
or EUV waves.
Now Extreme Ultraviolet (EUV) images obtained by
the {\it Atmospheric Imaging Assembly} (AIA) 
on board the {\it Solar Dynamics Observatory} (SDO)
provide a significantly improved view of 
these large-scale coronal propagating fronts (LCPFs).
Between April 2010 and January 2013, a total of 171 LCPFs have been 
identified through visual inspection of AIA images in the 193~\AA\ channel.
Here we focus on the 138 LCPFs that are seen 
to propagate across the solar disk, first studying how they are associated 
with flares, coronal mass ejections (CMEs) and type II radio bursts.
We measure the speed of the LCPF in
various directions until it is clearly altered by active regions or
coronal holes.  The highest speed is extracted for each LCPF.  
It is often considerably higher than 
EIT waves. 
We do not find a pattern where faster LCPFs decelerate and slow
LCPFs accelerate.
Furthermore, the speeds 
are not strongly correlated with 
the flare intensity or CME magnitude, nor do they show an
  association with type II bursts.
We do not find a good correlation either
between the speeds of LCPFs and CMEs in a subset of 
86 LCPFs observed by one or both of
the {\it Solar and Terrestrial Relations Observatory} (STEREO) spacecraft 
as limb events.

\end{abstract}




\section{Introduction}

The so-called EIT wave\footnotemark \citep{Moses97, Thompson98}
is a propagating intensity enhancement in coronal EUV lines, 
notably in Fe~{\sc xii} lines around 195~\AA\ at $\approx$1.5~MK.  
Its discovery by the {\it Extreme Ultraviolet Imaging Telescope}
\citep[EIT;][]{Boudine95} on board the {\it Solar and Heliospheric
  Observatory} (SOHO)
sparked renewed interest in global disturbances 
that had been associated with
solar flares.  The most typical example of such disturbances 
is the Moreton-Ramsey wave in H$\alpha$ images \citep{Moreton60}.
\cite{Uchida68} modeled Moreton-Ramsey waves 
as chromospheric intersections of flare-launched fast-mode MHD waves
in the corona, where the magnetoacoustic speed is 
an order of 1000~km~s$^{-1}$, comparable to the typical speed of 
Moreton-Ramsey waves.
The same fast-mode MHD waves may be 
responsible for metric type II radio bursts  
and filament oscillations that had been found even earlier.

\footnotetext{It is also referred to as the EUV wave to remove the
  dependency on the instrument.  For clarity, however, we simply use
  the EIT wave throughout the paper.}

EIT waves were initially thought to be the coronal counterparts of 
Moreton-Ramsey waves
\citep[e.g.,][]{Thompson99}.  
However, this idea started to be questioned.  
First, the typical speed of EIT waves
\citep[200\,--\,400~km~s$^{-1}$, see][]{Thompson09} is much lower than
that of Moreton-Ramsey waves.
A magnetoacoustic speed that matches the EIT wave initially implied 
a high plasma $\beta$ 
\citep{Wang00,Wu01}, not usually applicable in the low corona,
although later works \citep{Ofman02,Wu05,Schmidt10,Downs11} 
have shown that this is not always the case.
Second, EIT waves are more
intimately associated with 
coronal mass ejections (CMEs) than with flares
\citep{Biesecker02, Cliver05, PFChen06}, 
whereas the MHD waves responsible for Moreton-Ramsey waves 
are often attributed to flares as were in Uchida's original concept
in the pre-CME era.
Third, some of the reported EIT waves appear to represent stationary brightening \citep{Delannee99}.

One way of interpreting these properties of EIT waves
is that they are not fast-mode MHD waves, but that they are signatures
more directly related to CMEs.  They include current
shells \citep{Delannee99}, stretched CME loops \citep{PFChen02}, 
and reconnection of CMEs with quiet-Sun bipoles \citep{Attrill07}.
Furthermore, slow mode waves \citep{Wang09}
and solitons \citep{Wills-Davey07} were also proposed as
contributors to EIT waves.   

However, it is still more common to interpret EIT waves in terms real
waves, fast-mode MHD waves in particular.
In certain events observed in H$\alpha$, soft X-rays, 
He~{\sc i}~$\lambda$~10830 
and radio (at both microwave and metric wavelengths), the tracks of 
the EIT wave fronts are found to be consistent with flare-generated 
large-amplitude MHD waves that undergo deceleration 
\citep{Warmuth01, Warmuth04a, Vrsnak06}.
Other characteristics that support the wave nature of EIT waves include
the fronts that decelerate \citep{Long11a}, 
broaden and decrease in amplitude with time \citep{Veronig10}, and
deflect at coronal hole boundaries \citep{Gopalswamy09}.

It appears that the lack of a consensus view of the EIT wave is due
partly to the lack of its definition beyond 
``the outermost propagating intensity front reaching global scales'' 
\citep{Patsourakos12}.  Another major factor is
the 10\,--\,20 minute cadence of EIT, which likely failed to 
capture the early phase of many events, where the speed may have been higher.  
Indeed, higher speeds and decelerating profiles were obtained 
using 
data from the {\it Extreme Ultraviolet Imager}
\cite[EUVI;][]{Howard08, Wuelser04} on the 
{\it Solar Terrestrial Relations Observatory} (STEREO), 
which have slightly better cadence 
\citep[see][2011a]{Long08}.
Note, however, that the maximum speed found by Long et al. was only 
475 km s$^{-1}$ partly because of solar minimum conditions 
\citep[see][for an extensive list of EIT waves observed by EUVI during
2007\,--\,2009]{Nitta13}.  
Despite 
the problems arising from the low cadence, our knowledge of 
EIT waves 
is still built heavily on a limited set of EIT observations, 
especially on the list of events during March 1997\,--\,June 1998  
as compiled by \cite{Thompson09}.  

One remarkable departure in recent years,
however, is a growing recognition that 
EIT waves could contain
both wave and CME-related
components as first pointed out by \cite{Zhukov04}\footnotemark. 
Interestingly, this change of our perception
is driven by advanced 3d MHD numerical simulations for a few
events \citep[][2012]{Cohen09, Downs11} rather more than by 
observations.

\footnotetext{It is fair to mention that the co-existence of the wave
  and CME-related components was first shown by Chen et al. (2002) in
  their 2d MHD simulations of eruptions.  However, these authors
  associated EIT waves exclusively with the CME-stretched loops.}

Observationally, it is clear that a breakthrough in
understanding global coronal disturbances, EIT waves included,
should come with 
the {\it Atmospheric Imaging Assembly} 
\citep[AIA;][]{Lemen12} on board the {\it Solar Dynamics Observatory},
which provides uninterrupted full disk images with 
unprecedented high cadence, high sensitivity 
and broad temperature coverage.  
Already more than 30 papers using AIA data have been published 
that include discussions of EIT waves,  
delivering some promising results.
However, almost all the published papers on EIT waves using AIA data
have so far been  
case studies of one or a few events, 
not answering how special they are, 
reflecting their specific processes or conditions. 

In this paper, we take a complementary approach of studying
a large sample of events that look like EIT waves in AIA images, and focus on 
their speeds.  
This is motivated primarily by the recent attempt of  
distinguishing three populations of EIT waves on the basis of their kinematic
behaviors \citep{Warmuth11}.  According to Warmuth \& Mann, 
fast and decelerating EIT
waves may correspond to non-linear fast-mode MHD
waves (shocks), and those with moderate and constant speeds to linear fast-mode MHD
waves propagating at the local fast-mode speed.  The slowest EIT waves
may correspond to CME-related magnetic reconfiguration.
After studying similar phenomena in AIA data, however, it appears that 
EIT observations did not capture the full range of global disturbances
in the corona largely because of the poor cadence.  With AIA, we can 
more clearly observe how they propagate in short time intervals that used to 
be covered by less than a few EIT images.  It is likely that we now
deal with phenomena not adequately represented by ``EIT waves.''
Therefore in this paper we use the term ``Large-scale Coronal
Propagating Fronts'' (LCPFs) to refer to EIT-wave-like phenomena,
as temporally resolved in AIA data.
\S2 describes how we find LCPFs and maintain a catalog.
In \S3 we discuss how LCPFs are associated with other solar transient
phenomena.
We discuss the propagation speeds of LCPFs in \S4.  
The implications of this study are discussed in \S4.

\section{A Catalog of LCPFs}

We have found 171 LCPFs during April 2010\,--\,January 2013.  As of this
writing, the software to automatically detect and characterize EIT waves 
has not yet been implemented in AIA data pipeline 
as part of 
the Computer Vision
for the SDO \citep[cf.][]{Martens12}.  
Therefore we basically need to 
rely on visual inspection of images to find LCPFs.
As a first step, we review running difference images in  
211~\AA, 193~\AA, and 171~\AA\ channels 
at a sampling of every five minutes,
using a high-performance image-viewing tool called
Panorama \citep{Hurlburt12}.  
This exercise results in more than 200 candidates of LCPFs.  
In the absence of a universal and quantitative definition of EIT waves, 
our working definition of LCPFs is that they need to
exhibit an angular expanse of $\gtrsim$45$\arcdeg$ and 
to propagate at least 200~Mm away from the center of the associated eruption.
For each of the candidates, 
we measure the width and distance of the front 
in the last 193~\AA\ difference image on which
it can be traced.
We use EUVI 195~\AA\ images to measure the width and distance
of the front that comes from a region close to the limb and
propagates predominantly along the limb,  
leaving almost negligible signatures on disk.  For the period of
interest, limb events from Earth are typically 
viewed as disk events by one or
both of the STEREO spacecraft.  After dropping the candidates that do
not meet the above criteria, we are left with 171 LCPFs.  They are
listed in Table~1.

In order to study the global properties of LCPFs such as their
relations with CMEs, we need to analyze full-disk images.  
Considering our event sample size, it is not realistic 
at the moment to conduct an analysis in the full-resolution images of 4096$^{2}$
pixels.  Thus they are rebinned to 1024$^{2}$ pixels for the
present work.  Our experiences have shown that movies of coaligned
images are extremely useful for following the spatio-temporal
variations of dynamic phenomena on the Sun.
Therefore, for each LCPF, we make standard SolarSoft movies (Javascript and
MPEG) in AIA's seven EUV channels and three formats 
(intensity, running difference and base difference). 
Additionally, to benefit from stereoscopic views, 
two other sets of movies are made of AIA
193~\AA\ and STEREO-A or -B EUVI 195~\AA\ pairs. 
It is also important to follow LCPFs with respect to the development of the
associated flares as captured in soft X-ray light
curves.  To accommodate GOES soft X-ray light curves in the movies, 
we further shrink the
images to 768$^{2}$ pixels, although images of 
1024$^{2}$ pixels are used when measuring LCPFs (\S4).
LCPFs are usually found in 
running difference movies, but base difference movies are useful for 
isolating long-lasting dimming, which may correlate with 
the spatial extent of the CME \citep{Thompson00a}.  
Intensity movies, in which strong LCPFs are
visible, also help us locate coronal holes and active regions that 
deflect LCPFs.
The catalog of LCPFs with all these movies
are online at \url{http://aia.lmsal.com/AIA_Waves/index.html},
which may contain more events over time that are not used in the
present study.

Out of the 171 LCPFs, 138 are seen to propagate across the solar disk,
and 22 (mostly from regions close to the limb) to propagate predominantly over the
limb without clear fronts on disk.  The remaining 11 events do not
show a clear front in either way, 
although future image processing techniques may restore it. 
In the following sections, even
though a number of recent studies have examined propagations over
the limb \citep[e.g.,][]{Downs12, WeiLiu12}, our
emphasis here is on the first category, since many scientists may have
associated EIT waves with circular fronts propagating across the
disk. 

\section{Association of LCPFs with other observables}
 
We study the association of the 138 LCPFs that propagate across the
solar disk with 
solar flares, CMEs and type II radio bursts,
following a past study by \cite{Biesecker02} for the EIT waves that
were compiled by
\cite{Thompson09}.  In Table~1, they are given in the third, ninth and
tenth columns.
The GOES X-ray peak flux of the associated flares and the presence of
type II bursts are easily available from NOAA lists or SolarSoft
distributions.  
However, the same may not apply to CMEs because the online CME catalogs tend
to contain as many events as possible, some of which may be too
insignificant to qualify as CMEs.  
Rather than simply showing
whether the LCPF is associated with a CME as found close in time in a catalog,
we differentiate the significance of CMEs or outflows 
into four levels similar to the flare magnitude which is often 
referred to as X-class and so on.
By examining the available coronagraph
data (COR-1 and COR-2 on STEREO and LASCO on SOHO),
we introduce the following CME levels; (1) Weak
outflow that becomes invisible before the heliocentric distance 
of 5~R$_{\sun}$, (2) Narrow ($<$60$\arcdeg$) or 
slow ($<$500 km s$^{-1}$) outflow traceable beyond 5~R$_{\sun}$, 
but typically not
being reminiscent of the three-part CME structure, 
(3) Well-formed CME, fast and wide, with a flux rope or 
three-part structure, (4) Similar to
level 3, but very fast ($>$1500 km s$^{-1}$).  

Separating the CMEs into groups may not be done completely
  objectively, which is especially true for the distinction between 
  CMEs of levels 3 and 4, 
as many different threshold speeds could be considered.
Nevertheless, we consider a grouping like this to be important when we
understand how LCPFs arise in the overall picture of sudden energy
release in the solar corona.
\cite{Patsourakos09b} criticized
careless use of the term ``CME'' to include coronal structures 
that are not part of the flux rope.
Here we propose that how far it can travel may be another criterion
for a CME. 

Table~2 shows the breakdown into the flare class 
of the CME levels defined in this way and the presence/absence of a
type II burst.  Flares whose peak flux is $<$10$^{-6}$~W m$^{-2}$ are
labeled ``$<$C.''  
To our surprise, more than 1/3 of our LCPFs belong to CME level 1,
given the strong correlation of EIT waves with CMEs as widely accepted 
\citep{Biesecker02, Cliver05, PFChen06}.
Although we observe at least a minor outflow in the coronagraph data
of almost all of our LCPFs, 
many of them may not be real CMEs defined as ejections of
coronal magnetized plasma into the heliosphere.

\section{Speeds of LCPFs}

The speeds of the 138 LCPFs that propagate across the disk 
are measured in 193~\AA\ images (Figure~1), 
using a semi-automated scheme.
As the first step, we de-rotate the images to a reference time in order to
compensate for differential solar rotation,
using the standard SolarSoft routine \texttt{drot\_map.pro}.
The reference time is typically set to be 5\,--\,10~minutes before the onset of
the associated flare.
Then, as explained in \cite{WeiLiu10}, we follow a technique 
that is widely used for tracing waves \citep[see, for example,][who developed 
the Novel EIT Wave Machine Observing (NEMO) code]{Podladchikova05}.
First, the eruption center is identified as a pole. 
Here we make 24 equally spaced longitude sectors from the pole 
that are 15$\arcdeg$ wide.  
For each sector the intensity profile is obtained 
as a function of the distance along the longitude, 
by averaging pixels in the latitudinal direction.  
This corrects for the curvature of the solar
surface.  Such a profile is obtained 
at the cadence of 12~s (or 24~s after October 2010\footnotemark).  
The images with automatic exposure control are excluded  
to avoid spurious effects in difference images irrespective of 
which events we deal with.

\footnotetext{Then the automatic exposure control (AEC) was
  implemented on every other EUV image to reduce the effect of
  saturation in flare images.  To produce difference images unaffected
  by the spurious effects arising from different exposure times, we
  use only AEC-off images.}

We now have a 2d array of intensity as a function of both time
and distance from the eruption center. 
This should contain useful information
on the front such as the amplitude and width
\citep{Veronig10,WeiLiu10}.   However, extraction of
such information may not be straightforward on a large sample, because
it can be affected by local conditions specific to individual events.  
In this paper we instead concentrate on
the simplest quantity, namely the propagation speed,
which is measured on a distance-time plot.  In Figure~1, we show
distance-time plots   
for all the sectors as running difference images, on which LCPFs
appear as bright ridges.  We also include  
the normalized GOES 1\,--\,8~\AA\ soft X-ray flux 
as plotted on the same time axis.

For each sector, we calculate the speed and acceleration 
by fitting first- and second-order polynomials to the front edge of 
the most prominent ridge in the distance-time plot.  MPFIT 
\citep{Markwardt09} is used to calculate these parameters with error
bars, assuming a uniform uncertainty of 5~Mm in locating the front.
This is done only in the distance range until  
the propagation is clearly altered by active regions or coronal holes.  
Beyond this range, the technique may not make
sense because we are probably not tracing the same front. 
Furthermore, the technique may not work as well in data
whose temporal resolution is much worse than that of AIA 
\citep[e.g.,][]{Nitta13}. 
In intensity images (not shown), active regions are seen in the
sectors 1\,--\,3, 6, 7, 10, 11, 21, 23 and 24 and coronal holes 
in sectors 12\,--\,17, where the main ridge shows clear deceleration.
We examine the distance-time plot in each sector 
and select the one with the highest speed,
making sure not to select a sector in which the measurement is 
too susceptible to foreshortening toward the limb.
In this case sector 4 is selected, 
and the speed measured in that sector is registered as the speed
of this LCPF (see the sixth and seventh columns in Table~1).

Figure~2 shows distance-time plots from images in 171~\AA.
Generally, the fronts are harder to trace. 
Moreover, they appear as depression in several sectors
rather than as intensity enhancement.  
This may indicate heating at the
fronts \citep{Wills-Davey99, Schrijver11,WeiLiu12}, for example, 
from $\lesssim$1~MK to $\approx$1.5~MK.
In this example, the front
in sector~4 still appears as intensity enhancement.
The eighth column in Table~1 shows 
the appearance of the front in 171~\AA\ channel, summarized in the
last three columns in Table~2.  
In 171~\AA\ images, 
42\% of LCPFs
are not clearly seen (below the noise level) in the sector of the fastest
front at 193~\AA\ and another 42\% are seen as 
intensity depression.

Let us discuss the 22 LCPFs that propagate 
predominantly over the limb.  
We measure their speeds 
along the limb at the heliocentric distance of 1.15~R$_{\sun}$
\citep[see][for stereoscopic determination of the height of EIT waves]
{Patsourakos09a, Kienreich09}, as shown in Figure~3.  
In Figure~4, they are compared with the speeds measured in 
EUVI 195~\AA\ images of the same LCPFs in the same directions (north
or south) viewed as disk events.
The error bars for the speeds from AIA are based on the uniform
uncertainty of locating the front to 5~Mm.  
It is more difficult to measure the speeds of
faster LCPFs in EUVI data,
given the 2.5\,--\,5~m cadence. 
Therefore the error bars for the speeds
from EUVI are essentially proportional to the speeds with 
other corrections due to the cadence and the visibility
of the fronts, which may be still tentative.
Irrespective of the error bars, Figure~4 indicates that the two speeds
are not well correlated.
In many events, the AIA speed is higher than the EUVI speed, presumably
reflecting the difference in the cadence \citep[cf.][]{Long08,Long11a}.  
But other LCPFs show the opposite trend, suggesting that the limb view of
LCPFs after all reflect line-of-sight integration of the fronts whose
propagations may not only be in the north-south directions but also in 
other directions.  
Although it may be worthwhile to investigate reasons 
(e.g., projection effect) for the
discrepancy of the speeds in individual cases,
we generally argue that these events may not be
directly comparable to the LCPFs that are seen to propagate across
the solar disk in AIA images.

For the 138 LCPFs that are seen to propagate across the solar disk, 
we get $v_{{\tt mean}}$=644~km s$^{-1}$ and $v_{{\tt median}}$=607~km s$^{-1}$, which are
much higher than the typical speed of EIT waves 
\citep[200\,--\,400~km s$^{-1}$, see][]{Thompson09}.  
Figure~5 shows the speed vs acceleration of these LCPFs.  
Up to $v \approx$800~km~s$^{-1}$, there may be a weak trend of 
faster LCPFs decelerating as was the case for EIT waves
\citep{Warmuth11}.  However, in the full speed range, acceleration is
distributed more or less around zero.  This is partly because we fit
the distance vs time only until 
the propagation is clearly altered by active regions or coronal holes.  
Furthermore, there seems no difference in
the speed vs acceleration pattern when the LCPF is associated or not
associated with a type II burst.
Figure~6 shows the distribution of linear speed with respect to the
flare class, CME level (see the previous section), association with
type II bursts and appearance of the front at 171~\AA.  On average the
speed appears to be correlated with these observables.  Higher speeds
are seen in intense flares and energetic CMEs, and when type II
bursts are associated and the 171~\AA\ front appears in intensity
enhancement.  This is summarized in the average and median speeds for
each of the observables.
However, the distribution is broad and it is possible to
find high-speed LCPFs without the above properties.

Lastly, we compare the speeds of LCPFs and CMEs (or outflows)
observed by COR-1 in the heliocentric distance range of (1.5\,--\,4)R$_{\sun}$,
for a subset of 86 events for which the source region is 
located limbward of 60$\arcdeg$ longitude from
either or both STEREO spacecraft. 
The speeds measured in this view may be closer to the true CME speeds 
because they are less susceptible to projection and
visibility effects 
\citep[e.g.,][]{Burkepile04,Yashiro05}.
Figure 7(b) gives an example of the distance-time or height-time plot of
a CME along the cut as shown in Figure~7(a).  We make sure that the
cut passes both the LCPF as projected on the limb and the dominant
direction of CME propagation.
In Figure~8 we compare the speeds of 86 LCPFs and associated CMEs (or
outflows) as observed by COR-1 as limb events.  We assume the uncertainty of 25~Mm in
locating the CME front (Figure~7(b)).  
The two speeds do not seem to be strongly correlated, irrespective of
whether the LCPF is associated with a type II burst.

\section{Discussion}

We have found 171 LCPFs 
during April 2010\,--\,January 2013 by
manually inspecting AIA images, and made an online catalog of LCPFs.  This is
not meant to be a complete catalog, but we believe that most ``major'' LCPFs
have been included.  Some minor events discussed by 
\citet[][2012c, 2012d, 2013]{Zheng11} are not included in Table~1
because of our
working definition of LCPF that it should be $>$45$\arcdeg$ wide and   
be observed more than 200~Mm away from the eruption center.
The last column in Table~1 shows the references so far published to
discuss the individual LCPFs either directly or indirectly.  To date a
small number of relatively old events have been studied.

As captured in the movies included in the catalog, LCPFs have widely
different appearances, many of which are not circular, as was thought
to be ``typical'' for EIT waves \citep[see a review by][]{Wills-Davey09}.
Some LCPFs appear predominantly over the limb as seen in AIA images.
Here we do not discuss them because the speeds obtained by EUVI in
disk view are sometimes far from those measured by AIA along the limb,
although such limb events have been used in the context of the numerical
simulations that have helped us recognize the coexistence of wave and
CME-related components \citep[][2012]{Downs11}.  Some events close to
the limb may even
show more radial motions than lateral, marginally producing LCPFs
\citep[e.g., 2010 August 1 event, see][]{Schrijver11, WeiLiu11}.  
It is to be seen how these ``untypical'' events will be included 
in the automatic detection software being developed. 
Such software should make it easier to conduct ensemble studies of
LCPFs on more involved parameters.
This study may help validate the software.

It is widely accepted that EIT waves are more intimately associated
with CMEs than with flares \citep{Biesecker02, Cliver05, PFChen06}.
In many LCPFs, however, the associated CMEs are quite insignificant,
suggesting that the processes close to the Sun that allow fast lateral
expansion seems to be a key for certain LCPFs \citep{Patsourakos12}.
However, it is not clear whether such processes accompany all CMEs
and the height of their occurrences possibly determines LCPF
detection.  They are certainly not part of flare processes,
since for many LCPFs the associated flares 
are less intense than the GOES C-class.  It is
thus important to understand how the lateral expansion starts.  The present
study primarily deals with the global properties in 4$\times$4 rebinned full-disk
images, with the pixel resolution comparable to that of EIT.  On that
scale, we can easily miss out important changes within active regions that may
directly indicate lateral expansions.   Unlike EIT, however, 
we can go back to the full
resolution zooming into the active region and immediate neighborhood,
where, in combination with vector magnetic field data from 
the {\it Helioseismic and Magnetic Imager} \citep[HMI;][]{Scherrer12},
we may be able to observe the first 
signatures of the lateral expansion. 
In other words, this information may help us understand the range of variations
in how solar eruptions occur. 

Even though there were criticisms \citep[e.g.,][]{Attrill10} to the
idea of deflection of EIT waves at coronal holes \citep[e.g.,][]{Gopalswamy09},
AIA data show that such deflections or alterations of the
front are quite common at non-quiet-Sun magnetic field elements.
Movies included in our online catalog of LCPFs as described in \S2
immediately show us how LCPFs propagate outside quiet-Sun regions.  It
is likely that these deflections and subsequent propagations indicate
that they are fast-mode MHD waves, which at greater distances likely
represent freely propagating waves.  In this work the speeds of LCPFs
are measured before they encounter active regions or coronal holes and
get deflected.  It is in this early stage that LCPFs are poorly understood.
Models indicate that in the early development of LCPFs 
the waves are not easily separable from CME loops
\citep{Downs12}.  Observationally, we may see both components
\citep{Patsourakos12}.  Further progress requires more
detailed analysis of the kinematics 
\citep[including the amplitude and width
of the front, e.g.,][]{Veronig10} and thermodynamic properties,
and runs of MHD simulations in wider parameter space than what have
been done for relatively modest eruptions 
\citep[][2012]{Cohen09, Downs11}.

As in Figure~5, we fail to confirm distinct kinematic properties as
found by \citet{Warmuth11} in EIT and EUVI data that may indicate their
different origins.  This is partly because our measurements are
restricted to the early stages of LCPFs. In other words, 
in at least a number of cases, the apparent
deceleration of EIT waves probably resulted from deflection by coronal
holes or active regions at large distances.  In the past, several
examples of deceleration were found by comparing the trajectories of
EIT waves with those of faster waves observed in other wavelengths
such as H$\alpha$, soft X-rays, He~{\sc i}~$\lambda$~10830, and radio.
It is possible that EIT waves observed in other wavelengths represent
a subset of ``strong'' events, and that AIA has observed only a few of them.
Indeed, we know of only two Moreton-Ramsey waves
since SDO launch \citep[][White, 2012, personal
  communication]{Asai12}, but it is possible that we may find more
with more extensive search.  Now waves in other wavelengths can be
directly compared with LCPFs without extrapolation in time.  
Such comparisons will be useful for further clarifying the relation between
the wave and CME components in LCPFs in the early phase and probing
the origin of large-amplitude fast MHD waves.  The sharpness of the
front may also be useful for answering these questions, since in the past
``sharp'' waves accompanied EIT waves observed in multiple wavelengths
\citep{Thompson00b}.
The association of
LCPFs with Moreton-Ramsey waves and type II bursts may also depend on the angle of
the propagating front with respect to the solar surface that can
affect the downward pressure and the shock geometry important for 
particle acceleration
\citep{WeiLiu12}.

We find in Figure~6 some correlation between the speeds of LCPFs and
indicators for the magnitude of eruption or energy release.  However
the correlation is not strong.  There is an expectation
that the correlation could be made tighter if the magnitude of energy
release were more properly formulated such as using the X-ray fluence
rather than the peak flux\,--\,which may not be trivial especially for
small flares\,--\,, but it is more likely that the occurrence and
speed of LCPFs depend on the external conditions rather than the
energy release mechanism.  This is consistent with the fact that a
small number of active regions are very prolific  in LCPFs (see Table~1).

In this study we do not deal with LCPFs at greater distances beyond 
surrounding active regions and coronal holes.  It appears that such
LCPFs may be freely propagating MHD waves.  They may still play 
an important role in acceleration and transport of solar energetic
particles  \citep[e.g.,][]{Krucker99,Rouillard12} and in sympathetic
flares and eruptions \citep[e.g.,][]{Schrijver13}.  For these questions,
we need to analyze data in longer time ranges beyond individual LCPFs,
and the 360$\arcdeg$ view of the Sun made by combining AIA
and EUVI data \citep{Olmedo12} would be extremely useful.

Many papers have posed questions like ``What is the nature of
EIT waves?'' as if a single scheme could explain the phenomena in a
unified way even though different examples appear to have widely different properties.
It is more productive to characterize
individual LCPFs to understand when, where and how both wave
and CME components appear \citep{Schrijver11,Downs12}.  This study may help to put into
perspective the individual LCPFs that will be studied in detail.

\acknowledgments

This work was supported by the AIA contract NNG04EA00C
to LMSAL.  WL acknowledges NASA grant NNX11AO68G.  We thank the 
referee for useful suggestions for us to improve the manuscript.





\clearpage
\begin{figure}
\includegraphics[scale=.80]{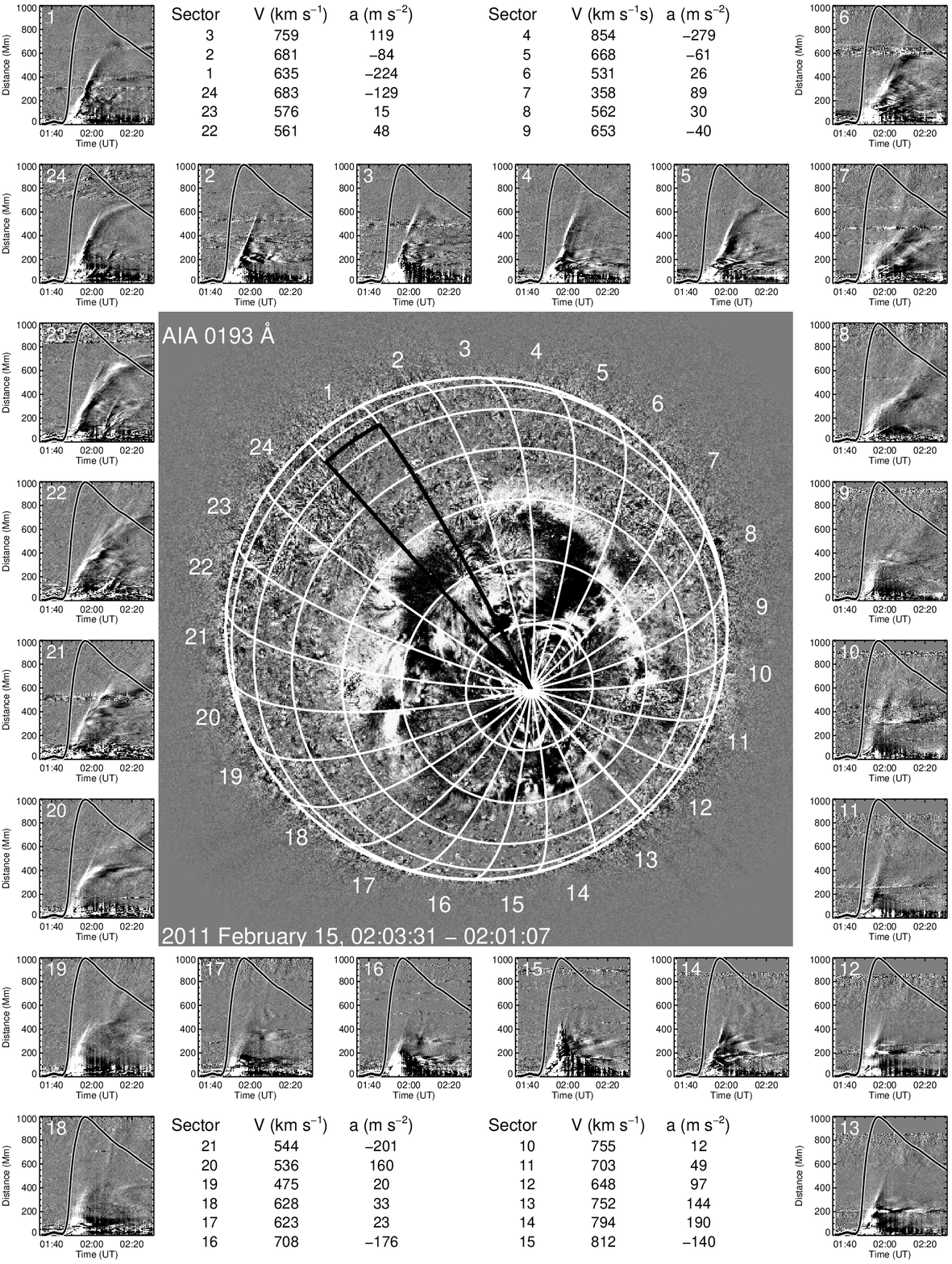}
\figcaption{This figure shows 24 running-difference distance-time
plots made from AIA images in 193~\AA\ channel 
for the 2011 February 15 LCPF.  The plots are made 
in equally spaced sectors that are bounded by two great circles 
passing through the flare, as indicated in the image placed in the
center.  The normalized GOES 1\,--\,8~\AA\ light curve is added to
each distance-time plot.
The distance is measured along the great circle.  The distance of
1000~Mm is encircled in black in sector~1.
The measured
linear speed and constant acceleration are also shown.}
\end{figure}

\clearpage
\begin{figure}
\includegraphics[scale=.80]{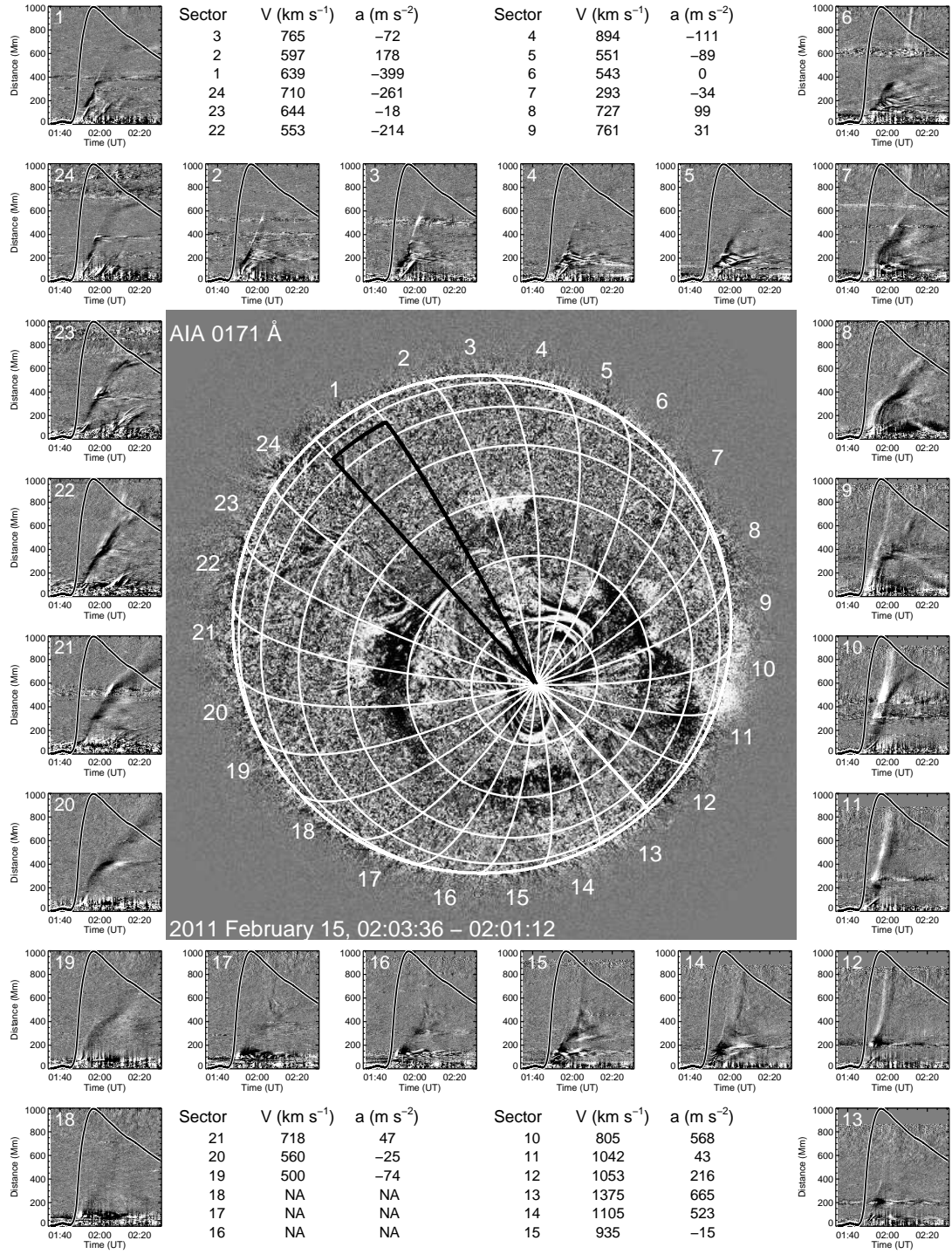}
\figcaption{Same as Figure~1 but in 171~\AA.}
\end{figure}

\clearpage
\begin{figure}
\plotone{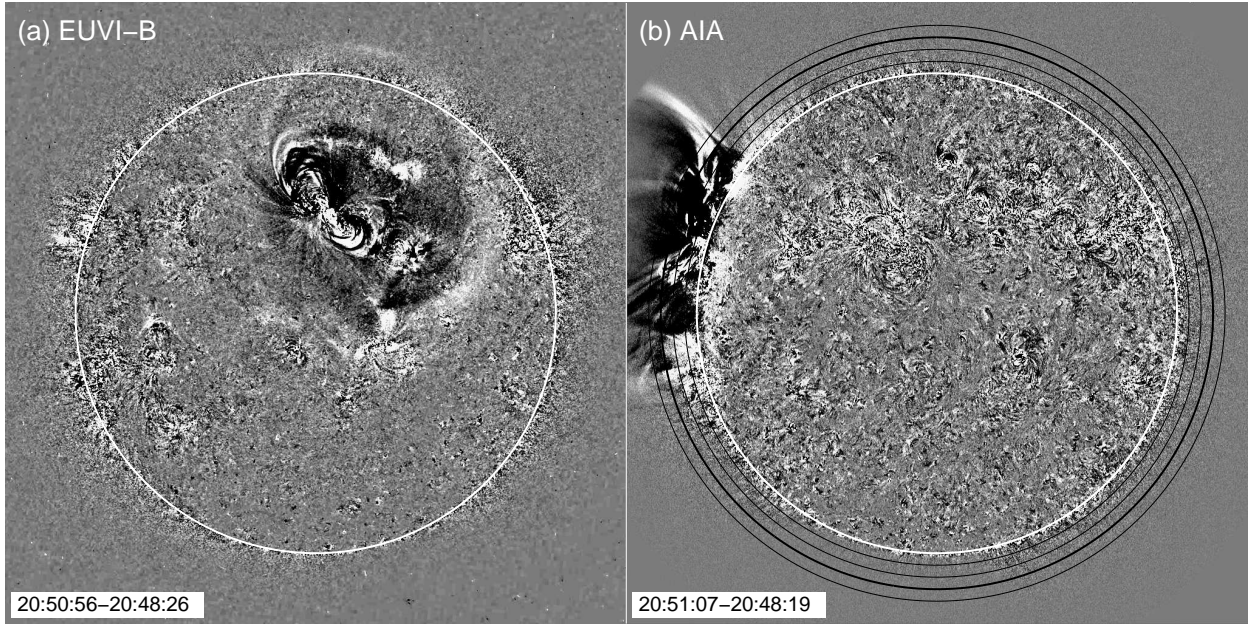} 
\figcaption{An example of LCPF seen predominantly along
  the limb with little presence on disk as seen by AIA.  Panel (b)
  shows an AIA 193~\AA\ difference image with black circles of heliocentric
  distances of 1.05R$_{\sun}$, 1.10R$_{\sun}$, 1.15R$_{\sun}$ and 
1.20R$_{\sun}$ in addition to the limb in white.  The circle of
1.15R$_{\sun}$ is shown in a thick line.  The same front is
seen by EUVI on STEREO-B as a disk event as shown in panel (a).}
\end{figure}

\clearpage
\begin{figure}
\plotone{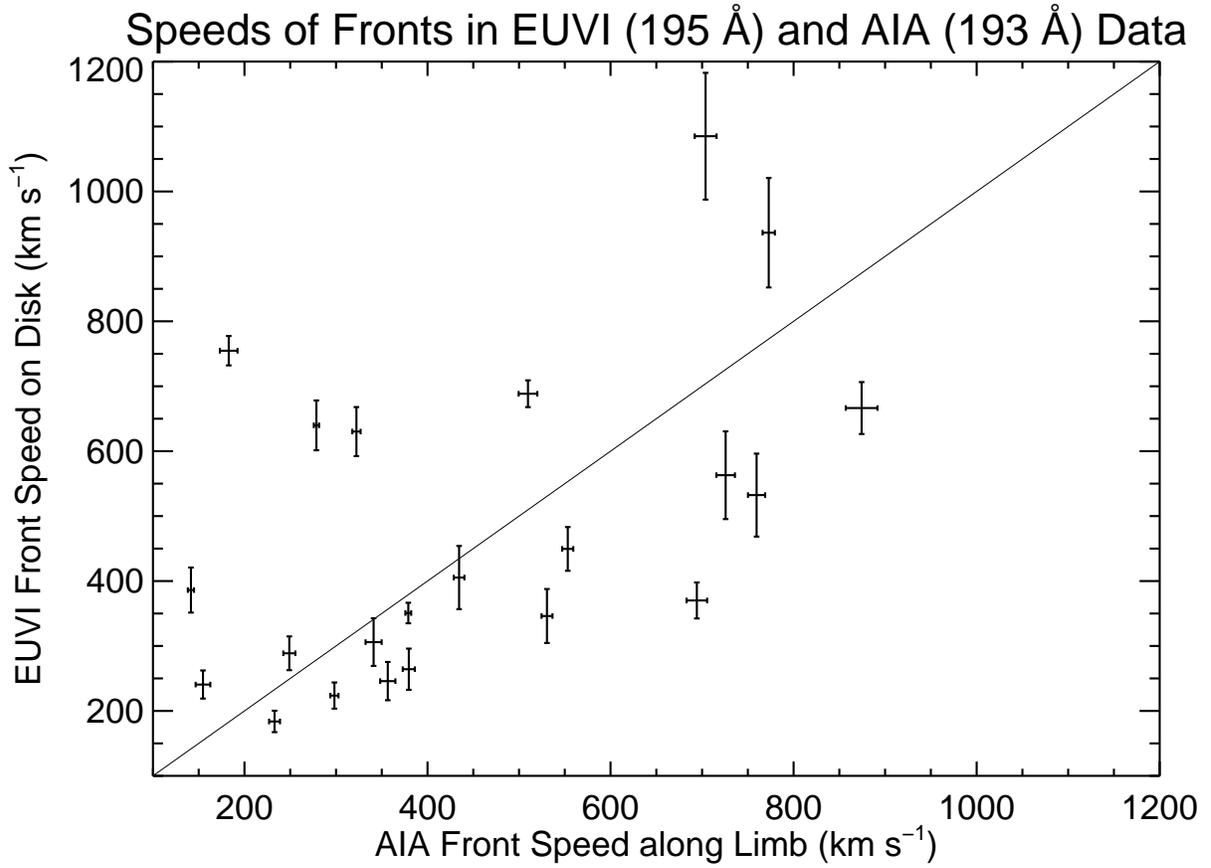} 
\figcaption{Comparison of the speeds of the same fronts measured along
the limb (at the heliocentric distance of 1.15~R$_{\sun}$) by AIA and
along great circles on disk by EUVI. The line gives $v_{{\tt AIA}} =
v_{{\tt EUVI}}$ as a visual guide. }
\end{figure}

\clearpage
\begin{figure}
\plotone{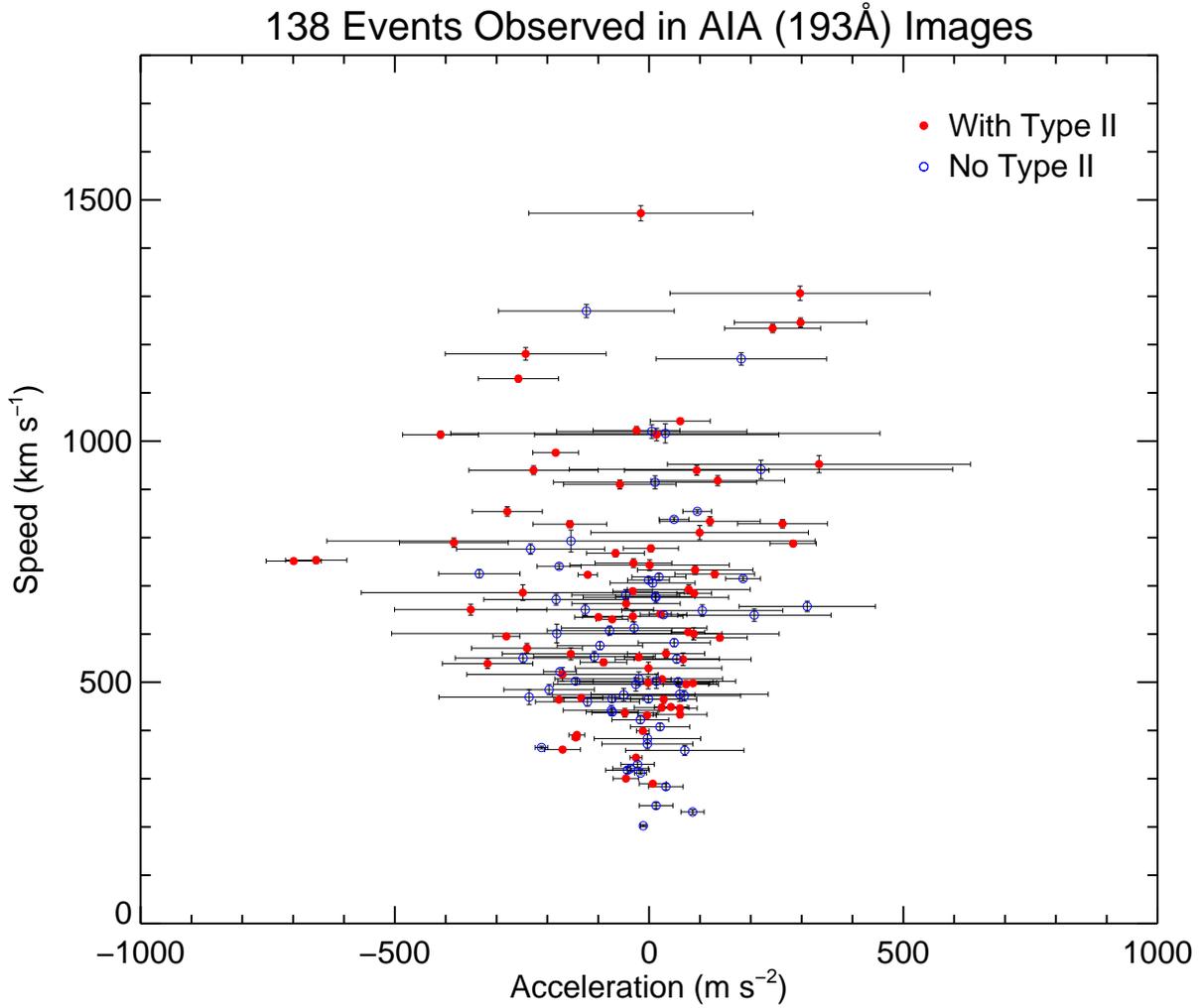} 
\figcaption{Linear speed of the propagating front in 193~\AA\ plotted against 
constant acceleration in 138 events that allow us to trace the front
on disk.  Data are plotted
separately for events with and without an associated type II radio burst.}
\end{figure}

\clearpage
\begin{figure}
\plotone{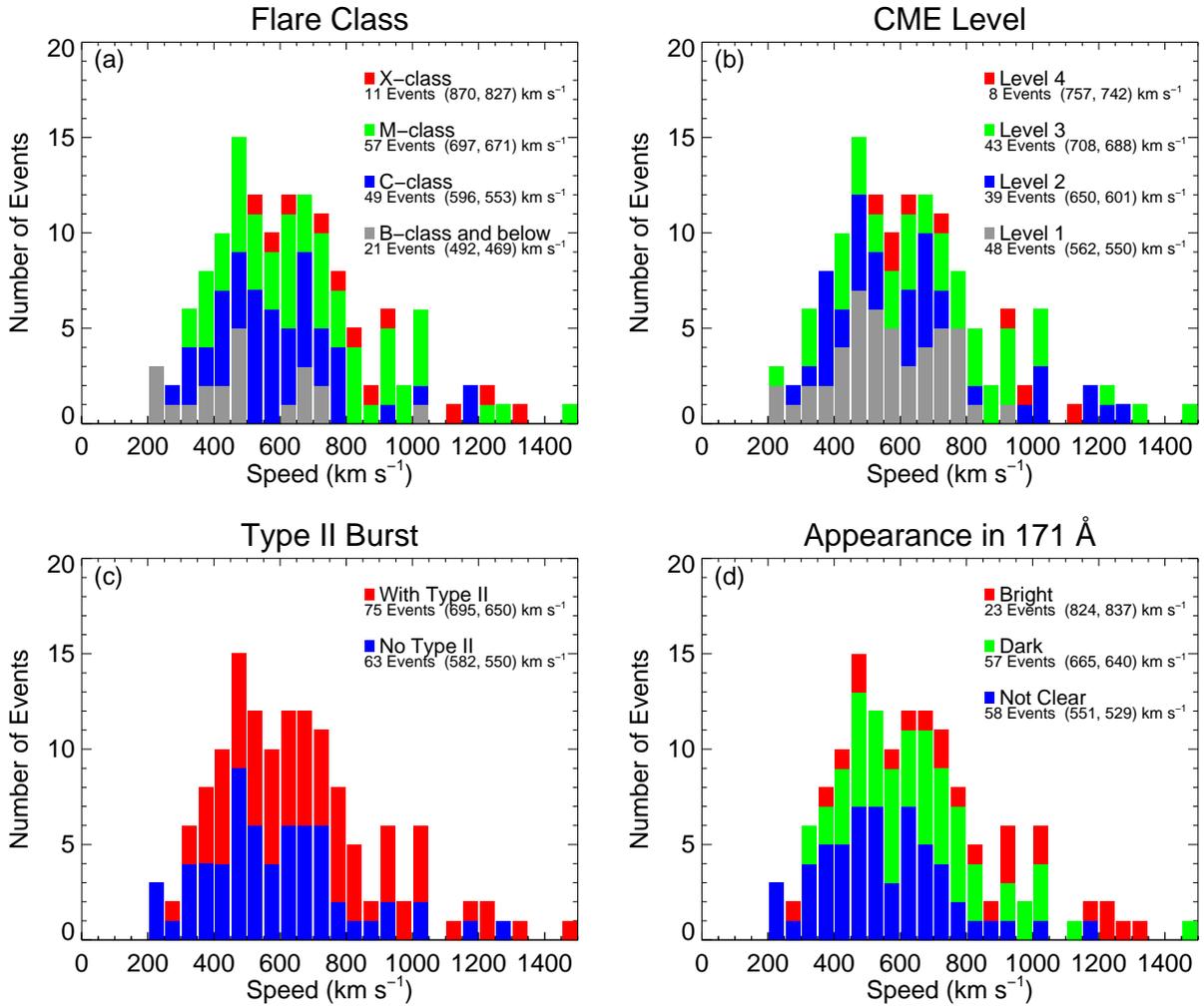} 
\figcaption{Histogram of the speed of LCPFs plotted
  for four observables.  Average and median speeds are shown for each
  category of the observables. }
\end{figure}

\clearpage
\begin{figure}
\plotone{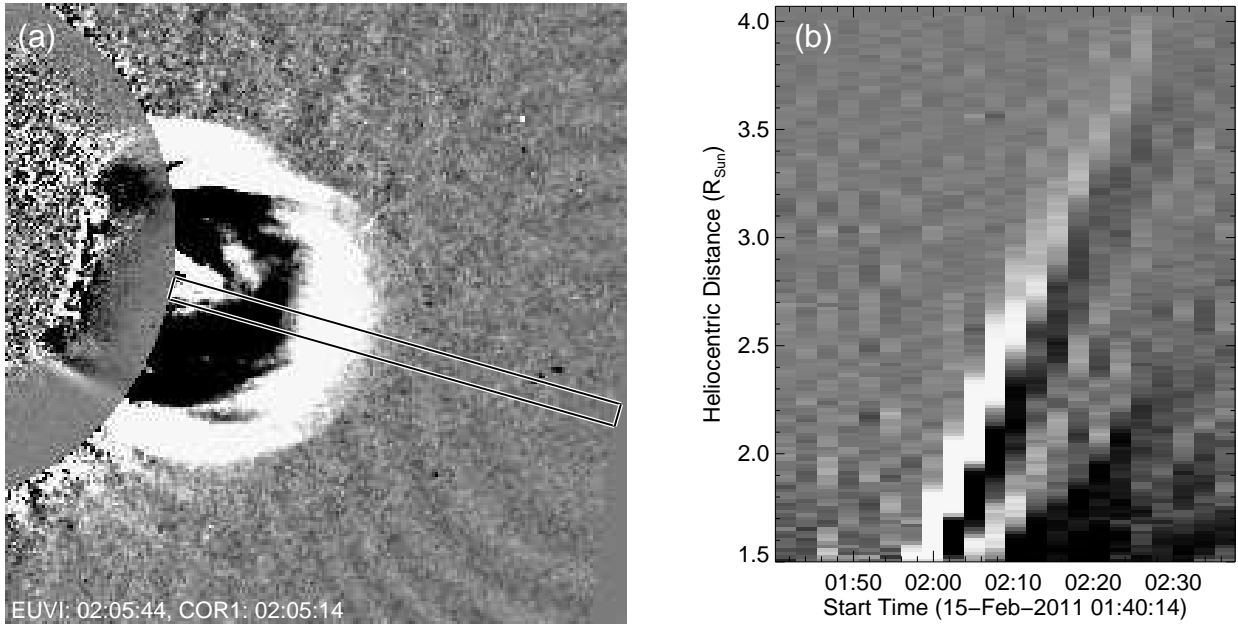} 
\figcaption{(a) Composite of EUVI and COR-1 difference images.  The
  distance-time plot of COR-1 data is shown in (b) for the cut indicated by a
  rectangle in (a). }
\end{figure}

\clearpage
\begin{figure}
\plotone{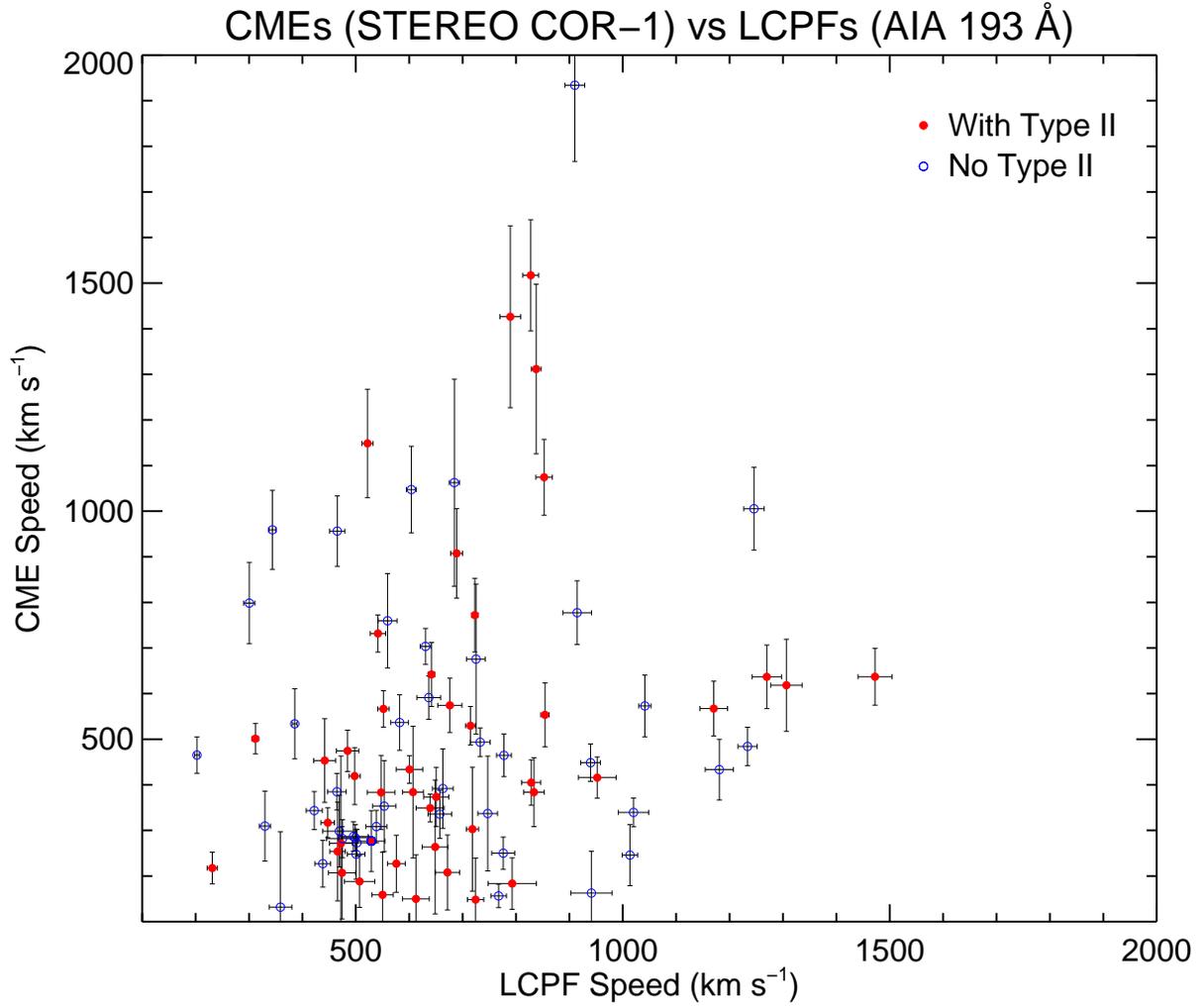} 
\figcaption{The CME speed as derived from COR-1 data plotted against
  the speed of LCPFs for events observed limbward of the longitude
  of 60$\arcdeg$ by STEREO. }
\end{figure}

\clearpage
\hoffset=-0.4in  

\onecolumn

\begin{deluxetable}{lllllcrrrcrrcc}
\tabletypesize{\scriptsize}

\tablecaption{The 171 LCPFs} 
\tablecolumns{14} \tablewidth{0pt}

\tablehead{\multicolumn{5}{c}{Flare} & \colhead{} & \multicolumn{3}{c}{LCPF} & 
\colhead{} & \multicolumn{4}{c}{Others} \\
\cline{1-5} \cline{7-9} \cline{11-14} \\
\colhead{} &           \colhead{Start} &     \colhead{GOES} &          \colhead{} & 
\colhead{} &           \colhead{} &          \colhead{Speed} &         \colhead{} & 
\colhead{} &    \colhead{} &          \colhead{CME} &           \colhead{Type II} & 
\colhead{STEREO} &  \colhead{} \\
\colhead{Date} &       \colhead{Time} &      \colhead{Class} &         \colhead{Location} & 
\colhead{AR} &         \colhead{} &          \colhead{(km s$^{-1}$)} & \colhead{Direction} & 
\colhead{171~\AA \tablenotemark{a}} & \colhead{} &          \colhead{Level} &         \colhead{Burst} & 
\colhead{Limb} &     \colhead{References\tablenotemark{b}} }

\startdata 
2010 Apr 8 &  02:30 & B3.7 & N24 E17  & 11060 & &  203 & S  & N & & 3 & N  & A   & 15 \\  
2010 May 4 &  06:30 & B1.3 & S26 E04  & 11066 & &  469 & E  & N & & 2 & N  & BA  & ... \\  
2010 May 8 &  20:04 & C2.4 & N40 W75  & 11069 & &  502 & S  & D & & 2 & N  & ... & ... \\  
2010 Jun 12 & 00:53 & M2.0 & N23 W43  & 11081 & &  386 & SE & D & & 2 & Y  & B   & 3, 11  \\  
2010 Jun 12 & 09:04 & C6.1 & N22 W52  & 11081 & &  391 & SE & N & & 2 & Y  & ... & ...  \\  
2010 Jun 13 & 05:30 & M1.0 & S25 W84  & 11079 & &  360 & NE & D & & 2 & Y  & ... & 6, 7, 11, 19, 21  \\  
2010 Jul 16 & 15:08 & A8 & S21 W20  & N-AR  & &  231 & NW & N & & 1 & N  & B   & ...  \\  
2010 Jul 27 & 08:46 & A6 & S17 W48  & N-AR  & &  474 & N  & B & & 1 & N  & B   & 2  \\  
2010 Aug 1  & 07:22 & C3.1 & N20 E36  & 11092 & &  312 & N  & D & & 3 & N  & A   & 16  \\  
2010 Aug 7  & 17:55 & M1.0 & N11 E34  & 11093 & &  631 & S  & B & & 3 & Y  & A   & ...  \\  
2010 Aug 14 & 09:38 & C4.4 & N14 W54  & 11093 & &  431 & S  & D & & 3 & Y  & ... & 18  \\  
2010 Aug 18 & 04:45 & C4.5 & N14 W105 & 11093 & & Limb & ... & ... & & ... & ... & ... & ...  \\  
2010 Sep 8 &  23:05 & C3.3 & N20 W90  & 11105 & & Limb & ... & ... & & ... & ... & ... & 9, 17  \\  
2010 Oct 16 & 19:07 & M2.9 & S20 W26  & 11112 & & 1013 & NW & D & & 2 & Y  & B   &  13 \\
2010 Nov 03 & 05:54 & C3.8 & S20 E98  & 11121 & & ...  & ... & ... & & ... & ... & ... & ... \\
2010 Nov 03 & 12:43 & C3.4 & S20 E97  & 11121 & & ...  & ... & ... & & ... & ... & ... & ... \\
2010 Nov 11 & 18:52 & B9.4 & N12 E28  & 11124 & & 359  &  S & N & & 1 & N  & A   &  27 \\
2010 Nov 11 & 19:40 & C1.3 & N13 E27  & 11124 & & 466  & SE & D & & 1 & N  & A   &  27 \\
2010 Nov 11 & 20:47 & B9.0 & N13 E26  & 11124 & & 472 & SE  & D & & 1 & N  & A   &  27 \\
2010 Nov 15 & 14:36 & B7.6 & S22 W44  & 11123 & & 365 & NE  & N & & 2 & N  & ... & ... \\
2010 Dec 15 & 14:31 & B2.2 & N20 W55  & 11134 & & 651 & E   & N & & 2 & Y  & ... & ... \\
2010 Dec 31 & 04:18 & C1.3 & N13 W57  & 11138 & & 446 & NE  & D & & 2 & Y  & ... & ... \\
2011 Jan 27 & 08:40 & B6.6 & N13 W82  & 11149 & & ...  & ... & ... & & ... & ... & ... & ... \\
2011 Jan 27 & 11:53 & C1.2 & N13 W85  & 11149 & & 399 & SE  & B & & 2 & Y  & ... & 5 \\
2011 Jan 28 & 00:44 & M1.3 & N13 W90  & 11149 & & 464 & E   & D & & 3 & Y  & ... & ... \\
2011 Feb 11 & 21:31 & B8.1 & N11 E105 & 11161 & & Limb & ... & ... & & ... & ... & ... & ...  \\
2011 Feb 13 & 17:28 & M6.6 & S20 E04  & 11158 & & 498 & E   & B & & 2 & Y  & BA  & 8 \\
2011 Feb 14 & 02:35 & C1.6 & S21 E04  & 11158 & & 501 & N   & D & & 1 & N  & BA  & ... \\
2011 Feb 14 & 04:29 & C8.3 & S20 W01  & 11158 & & 501 & N   & N & & 1 & N  & BA  & ... \\
2011 Feb 14 & 06:51 & C6.6 & S20 W01  & 11158 & & 657 & NE  & D & & 1 & N  & BA  & ... \\
2011 Feb 14 & 12:41 & C9.4 & S20 W02  & 11158 & & 663 & NE  & D & & 2 & Y  & BA  & 8  \\
2011 Feb 14 & 17:20 & M2.2 & S20 W05  & 11158 & & 829 & NE  & D & & 2 & Y  & BA  & ... \\
2011 Feb 14 & 19:23 & C6.6 & S20 W05  & 11158 & & 639 & NE  & D & & 1 & N  & BA  & ... \\
2011 Feb 15 & 00:31 & C2.7 & S20 W07 & 11158  & & 651 & NE  & D & & 1 & N  & BA  & ... \\
2011 Feb 15 & 01:44 & X2.2 & S20 W12 & 11158  & & 854 & N   & D & & 3 & Y  & BA  & 20, 22 \\
2011 Feb 15 & 03:05 & $<$C1 & S21 W08 & 11158 & & 613 & NE  & N & & 1 & N  & BA  & ... \\
2011 Feb 15 & 04:27 & C4.8 & S21 W09 & 11158  & & 582 & NE  & D & & 1 & N  & BA  & ... \\
2011 Feb 15 & 05:07 & B1.3 & N22 E47 & 11161 & & 284 & S   & B & & 1 & N  & ... & ... \\
2011 Feb 15 & 07:11 & B2.7 & N21 E46 & 11161 & & 318 & S   & D & & 1 & N  & ... & ... \\
2011 Feb 15 & 14:32 & C4.8 & S20 W16 & 11158  & & 555 & N   & D & & 1 & N  & BA  & ... \\
2011 Feb 16 & 14:19 & M1.6 & S22 W30 & 11158  & & 496 & N   & N & & 1 & Y  & ... & 10, 26 \\
2011 Feb 17 & 21:30 & C1.1 & S18 W45 & 11158  & & 571 & N   & B & & 1 & Y  & ... & ... \\
2011 Feb 24 & 07:23 & M3.5 & N16 E88 & 11163  & & 433 & W   & D & & 3 & Y  & ... & ... \\
2011 Mar 7 & 13:45 & M1.9 & N12 E21 & 11166   & & 300 & N   & N & & 3 & Y  & BA  & 8  \\
2011 Mar 7 & 19:43 & M3.7 & N30 W47 & 11164   & & 506 & S   & D & & 4 & Y  & ... & ... \\
2011 Mar 8 & 03:37 & M1.5 & S21 E72 & 11171   & & 449 & NW  & B & & 3 & Y  & ... & ... \\
2011 Mar 8 & 18:50 & $<$B1 & N10 W01 & 11166  & & 442 & W   & D & & 1 & N  & BA  & ... \\
2011 Mar 8 & 19:46 & M1.4 & S19 W90 & 11165   & & Limb & ... & ... & & ... & ... & ... & ...  \\
2011 Mar 12 & 04:33 & M1.3 & N05 W36 & 11166  & & 753 & S   & D & & 1 & Y  & ... & ... \\
2011 Mar 12 & 15:19 & C9.6 & N05 W39 & 11166  & & 751 & S   & N & & 1 & Y  & ... & ... \\
2011 Mar 24 & 12:01 & M1.0 & S15 E44 & 11176  & & 712 & NW  & D & & 1 & N  & ... & ... \\
2011 Mar 25 & 23:08 & M1.0 & S12 E26 & 11176  & & 834 & NW  & D & & 1 & Y  &   B &  12 \\
2011 Mar 27 & 05:00 & A8 & N17 E101 & N-AR   & & Limb & ... & ... & & ... & ... & ... & ...  \\
2011 May 9 & 20:42 & C5.4 & N20 E91 & N-AR    & & Limb & ... & ... & & ... & ... & ... & ...  \\
2011 May 11 & 02:23 & B8.1 & N18 W52 & N-AR   & & 467  & SE & D & & 3 & Y  & ... & ... \\
2011 May 15 & 23:25 & C4.8 & N11 W47 & 11208  & & 706  & S  & N & & 2 & N  & ... & ... \\
2011 May 29 & 10:08 & M1.4 & S22 E65 & 11226  & & 810  & NW & N & & 3 & Y  & ... & ... \\
2011 May 29 & 20:11 & C8.7 & S18 E68 & 11227  & & Limb & ... & ... & & ... & ... & ... & ...  \\
2011 May 30 & 10:48 & C2.8 & S18 E60 & 11227  & & 290  & NW & N & & 2 & Y  & ... &  23 \\
2011 Jun 1 & 02:37 & C2.6 & S18 E37 & 11228   & & 548  & NW & D & & 1 & N  & ... & ... \\
2011 Jun 1 & 16:50 & C2.9 & S19 E19 & 11226   & & 408  & NW & D & & 1 & N  & ... & ...\\
2011 Jun 2 & 07:38 & C1.4 & S19 E20 & 11227   & & 465  & NW & D & & 3 & N  & B   & ...\\
2011 Jun 7 & 06:16 & M2.5 & S21 W54 & 11226   & & 976  & E  & D & & 4 & Y  & ... & 4, 14 \\
2011 Jul 3 & 00:01 & B9.5 & N14 W24 & 11244   & & 685  & S  & D & & 2 & Y  & BA  & ...\\
2011 Jul 11 & 10:37 & C2.6 & S17 E06 & 11249  & & 601  & SW & N & & 2 & Y  & BA  & ...\\
2011 Jul 30 & 02:06 & M9.3 & N15 E35 & 11261  & & 383  & S  & N & & 1 & N  & ... & ...\\
2011 Aug 2 & 06:00 & M1.4 & N15 W14 & 11261   & & 560  & S  & D & & 3 & Y  & BA  & ...\\
2011 Aug 3 & 13:17 & M6.0 & N17 W30 & 11261   & & 604  & S  & D & & 3 & Y  & A   & ...\\
2011 Aug 4 & 03:41 & M9.3 & N17 W37 & 11261   & & 910  & S  & B & & 4 & Y  & A   & ...\\
2011 Aug 8 & 18:00 & M3.5 & N16 W63 & 11263   & & 559  & SE & D & & 4 & Y  & ... & ...\\
2011 Aug 9 & 07:48 & X6.9 & N16 W72 & 11263   & & 743  & SE & D & & 4 & Y  & ... & 1, 24 \\
2011 Aug 10 & 15:46 & A1 & N20 E55 & N-AR   & & 686  & S  & N & & 2 & Y  & ... & ...\\
2011 Aug 11 & 09:34 & C6.2 & N16 W102 & 11263 & & Limb & ... & ... & & ... & ... & ... & ...  \\
2011 Sep 4 & 04:36 & C9.0 & S20 W108 & 11284  & & Limb & ... & ... & & ... & ... & ... & ...  \\
2011 Sep 6 & 01:35 & M5.3 & N14 W07 & 11283   & & 1041 & NW & B & & 3 & Y  & BA  & ...\\
2011 Sep 6 & 22:12 & X2.1 & N14 W18 & 11283   & &  1246 & N & B & & 3 & Y  & BA  & ...\\
2011 Sep 7 & 22:32 & X1.8 & N14 W28 & 11283   & &  1307 & N & B & & 3 & Y  & A   & ...\\
2011 Sep 8 & 15:32 & M6.7 & N14 W40 & 11283   & &  649 & NE & N & & 1 & N  & A   & ...\\
2011 Sep 9 & 06:01 & M2.7 & N16 W47 & 11283  & & ... & ... & ... & & ... & ... & ... & ...  \\
2011 Sep 22 & 10:29 & X1.4 & N12 E81 & 11302  & &  595 & W  & D & & 4 & Y  & ... & ...\\
2011 Sep 23 & 12:10 & C3.2 & N28 W52 & 11296  & & ... & ... & ... & & ... & ... & ... & ...  \\
2011 Sep 23 & 23:48 & M1.9 & N12 E65 & 11302  & &  516 & SW & N & & 2 & Y  & ... & ...\\
2011 Sep 24 & 09:32 & X1.9 & N12 E62 & 11302  & & 1129 & SW & D & & 4 & Y  & ... & 25 \\
2011 Sep 24 & 12:33 & M7.1 & N15 E60 & 11302  & &  640 & SW & D & & 4 & N  & ... & ...\\
2011 Sep 24 & 19:09 & M3.0 & N15 E60 & 11302  & & 918  & NW & B & & 3 & Y  & ... & ...\\
2011 Sep 24 & 21:23 & M1.2 & S28 W65 & 11303  & & 460  & N  & N & & 1 & N  & ... & ...\\
2011 Sep 24 & 23:45 & M1.0 & S28 W66 & 11303  & & 601 & NE  & N & & 2 & N  & ... & ...\\
2011 Sep 25 & 02:27 & M4.4 & S28 W68 & 11303  & & 372 & N   & N & & 2 & N  & ... & ...\\
2011 Sep 25 & 04:31 & M7.4 & N12 E50 & 11302  & & 740 & S   & N & & 3 & N  & ... & ...\\
2011 Sep 25 & 09:25 & M1.5 & S28 W72 & 11303  & & 321 & N   & N & & 1 & N  & ... & ...\\
2011 Sep 25 & 15:26 & M3.7 & N13 E43 & 11302  & & 678 & SW  & D & & 2 & N  & ... & ...\\
2011 Sep 26 & 14:37 & M2.6 & N14 E30 & 11302  & & 671 & S   & N & & 1 & N  & B   & ...\\
2011 Sep 27 & 20:34 & C6.4 & N14 E09 & 11302  & & 550 & S   & N & & 1 & N  & BA  & ...\\
2011 Sep 29 & 12:07 & C2.7 & N10 W11 & 11302  & & ... & ... & ... & & ... & ... & ... & ...  \\
2011 Sep 30 & 02:46 & C1.0 & N14 W32 & 11302  & & 576 & S   & D & & 1 & N  & A   & 28 \\
2011 Sep 30 & 03:47 & C7.7 & N10 E10 & 11305  & & 793 & S   & N & & 1 & N  & BA  & ...\\
2011 Oct 1 & 09:44 & M1.2 & N10 W06 & 11305   & & 733 & S   & N & & 3 & Y  & BA  & ...\\
2011 Oct 1 & 20:00 & ... & N24 E118 & ...     & & ... & ... & ... & & ... & ... & ... & ...  \\
2011 Oct 2 & 00:37 & M3.9 & N10 W14 & 11305   & & 715 & S   & D & & 3 & N &  BA  & ...\\
2011 Oct 2 & 21:41 & C7.6 & N10 W25 & 11305   & & 607 & S   & D & & 2 & N &  BA  & ...\\
2011 Oct 3 & 02:34 & C2.1 & S13 W62 & 11302   & & Limb & ... & ... & & ... & ... & ... & ...  \\ 
2011 Oct 10 & 14:30 & C4.5 & S13 E03 & 11313  & & 941  & W  & D & & 1 & N &  BA  & ...\\
2011 Oct 20 & 03:10 & M1.6 & N20 W95 & 11318  & & Limb & ... & ... & & ... & ... & ... & ...  \\
2011 Oct 21 & 12:53 & M1.3 & N06 W80 & 11319  & & 592 &  E  & N & & 3 & Y & ...  & ...\\
2011 Nov 9 & 13:04 & M1.1 & N20 E30 & 11342   & & 642 & NW  & N & & 3 & Y & B    & ...\\
2011 Nov 14 & 09:18 & C5.2 & N22 W63 & 11348 & & 1014 & E   & N & & 2 & Y & ...  & ...\\
2011 Nov 15 & 00:07 & $<$B1 & N08 E30 & 11347 & &  724 & S   & D & & 1 & Y & B    & ...\\
2011 Nov 24 & 23:57 & C1.5 & S18 W20 & 11354 & &  330 & SE  & N & & 2 & N & A    & ...\\
2011 Nov 25 & 21:49 & C2.4 & N17 W62 & 11359 & &  692 & SE  & N & & 1 & Y & ...  & ...\\
2011 Dec 13 & 03:08 & B2.3 & S17 E12 & 11374 & &  719 & W   & D & & 1 & N & BA   & ...\\
2011 Dec 22 & 01:56 & C5.4 & S19 W18 & 11381 & &  448 & N   & N & & 2 & Y & A    & ...\\
2011 Dec 25 & 08:49 & C5.5 & S21 W20 & 11387 & &  776 & E   & D & & 1 & N & A    & ...\\
2011 Dec 25 & 18:11 & M4.0 & S22 W26 & 11387 & &  940 & S   & B & & 3 & N & A    & ...\\
2011 Dec 25 & 20:23 & C7.7 & S21 W24 & 11387 & &  725 & E   & D & & 2 & N & A    & ...\\
2011 Dec 26 & 02:13 & M1.5 & S21 W33 & 11387 & &  854 & N   & N & & 3 & N & A    & ...\\
2011 Dec 26 & 11:16 & C5.7 & N18 W02 & 11384 & &  676 & NE  & B & & 3 & N & BA   & ...\\
2012 Jan 23 & 03:38 & M8.7 & N28 W21 & 11402 & &  837 & E   & B & & 3 & N & A    & ...\\
2012 Jan 27 & 17:37 & X1.7 & N27 W71 & 11402 & &  635 & E   & N & & 3 & Y & ...  & ...\\
2012 Mar 2 & 17:29 & M3.3 & N18 E86 & 11429   & & Limb & ... & ... & & ... & ... & ... & ...  \\ 
2012 Mar 4 & 10:29 & M2.0 & N17 E64 & 11429  & & 1016 & SW  & D & & 3 & N & ...  & ...\\
2012 Mar 5 & 03:25 & X1.1 & N17 E56 & 11429  & & 915  & SW  & N & & 3 & N & B    & ...\\
2012 Mar 7 & 00:02 & X5.4 & N18 E31 & 11429  & & 828  & W   & N & & 3 & Y & B    & ...\\
2012 Mar 7 & 01:05 & X1.3 & N18 E31 & 11429  & & 789  & W   & N & & 3 & Y & B   & ...\\
2012 Mar 9 & 03:22 & M6.3 & N17 E02 & 11429  & & 689  & W   & D & & 3 & Y & BA  & ...\\
2012 Mar 10 & 17:15 & M8.4 & N16 W24 & 11429 & & 522  & SE  & N & & 3 & N & A  & ...\\
2012 Mar 13 & 17:12 & M7.9 & N18 W61 & 11429 & & 1022  & E  & B & & 3 & Y & ...  & ...\\
2012 Mar 14 & 15:08 & M2.8 & N14 E06 & 11432 & & 485  & S   & N & & 2 & N & BA  & ...\\
2012 Mar 17 & 20:32 & M1.3 & S20 W25 & 11434 & & 548  & NE  & D & & 1 & Y & A   & ...\\
2012 Mar 26 & 22:40 & ... & N18 E125 & ...  & & Limb & ... & ... & & ... & ... & ... & ...  \\
2012 Apr 5 & 20:49 & C1.5 & N17 W32 & 11450  & & 552 & NE   & N & & 3 & Y & A & ... \\
2012 Apr 9 & 12:12 & C3.9 & N16 W63 & 11451  & & Limb & ... & ... & & ... & ... & ... & ...  \\
2012 Apr 23 & 17:38 & C2.0 & N14 W17 & 11461 & &  723 & SE  & B & & 1 & Y & A & ... \\
2012 Apr 24 & 07:38 & C3.7 & N12 E80 & 11467 & & 787 & SW   & D & & 3 & Y & ... & ... \\
2012 May 17 & 01:25 & M5.1 & N11 W76 & 11476 & & 939 & E    & D & & 3 & Y & ... & ... \\
2012 Jun 3 & 17:48 & M3.3 & N17 E38 & 11496 & & 1472 & W    & D & & 3 & Y &  B  & ... \\
2012 Jun 6 & 19:54 & M2.1 & S18 W05 & 11494 & & 1234 & S    & B & & 2 & Y &  A  & ... \\
2012 Jun 23 & 07:02 & C2.7 & N17 W102 & N-AR & & Limb & ... & ... & & ... & ... & ... & ...  \\
2012 Jul 2 & 10:43 & M5.6 & S18 E05 & 11515 & & 1234 & S    & B & & 2 & Y &  B & ... \\
2012 Jul 2 & 19:59 & M3.8 & S17 E03 & 11515 & & 1270 & S    & B & & 2 & N &  B & ... \\
2012 Jul 6 & 23:01 & X1.1 & S14 W59 & 11515  & & Limb & ... & ... & & ... & ... & ... & ...  \\
2012 Jul 8 & 16:23 & M6.9 & S14 W85 & 11515  & & Limb & ... & ... & & ... & ... & ... & ...  \\
2012 Jul 12 & 15:37 & X1.4 & S17 W02 & 11520 & & 542 & SW   & N & & 3 & Y &  BA & ... \\
2012 Jul 31 & 10:46 & C5.7 & N18 E64 & 11535     & & ... & ... & ... & & ... & ... & ... & ...  \\
2012 Aug 13 & 12:33 & C2.8 & N23 W04 & 11543 & & 1181 & NW  & N & & 2 & Y &  BA & ... \\
2012 Aug 14 & 00:23 & C3.5 & N23 W11 & 11543 & & 1170 & W   & B & & 2 & N &  A  & ... \\
2012 Aug 15 & 03:37 & B8.5 & N23 W25 & 11543 & & 1020 & N   & D & & 2 & N &  A  & ... \\
2012 Aug 16 & 12:41 & C3.6 & N23 W40 & 11543     & & ... & ... & ... & & ... & ... & ... & ...  \\
2012 Aug 17 & 22:23 & B5.9 & N23 W62 & 11543     & & ... & ... & ... & & ... & ... & ... & ...  \\
2012 Aug 31 & 19:45 & C8.4 & S27 E47 & 11562     & & ... & ... & ... & & ... & ... & ... & ...  \\
2012 Sep 15 & 22:23 & B9.6 & N26 W78 & 11566 & & 437 & E    & N & & 1 & Y & ... & ... \\
2012 Sep 16 & 22:05 & A5 & N18 W70 & N-AR & & 244 & E    & N & & 1 & N & ... & ... \\
2012 Sep 25 & 04:24 & C3.6 & N09 E20 & 11577 & & 465 & SE   & N & & 1 & Y & B   & ... \\
2012 Sep 27 & 22:36 & C3.7 & N09 W34 & 11575 & & 344 & SE   & N & & 3 & Y & A   & ... \\
2012 Nov 8 & 02:08 & M1.7 & N14 E82 & 11611  & & Limb & ... & ... & & ... & ... & ... & ...  \\
2012 Nov 10 & 04:22 & C2.0 & S25 E10 & 11608 & & 637 & E    & N & & 2 & Y & B   & ... \\
2012 Nov 18 & 03:55 & C5.7 & N08 W22 & 11615 & & 496 & SW   & N & & 2 & N & A   & ... \\
2012 Nov 20 & 19:21 & M1.6 & N05 E16 & 11618 & & 422 & SE   & N & & 1 & N & B   & ... \\
2012 Nov 21 & 06:45 & M1.4 & N05 E11 & 11618 & & 499 & S    & D & & 2 & Y & B   & ... \\
2012 Nov 21 & 15:10 & M3.5 & N05 E06 & 11618 & & 778 & E    & B & & 3 & Y & B & ... \\
2012 Nov 24 & 13:33 & C3.3 & N08 W29 & 11618 & & 539 & S    & N & & 1 & Y & A & ... \\
2012 Dec 5 & 00:12 & C1.7 & N17 E76 & 11628 & &  681 & SW   & N & & 2 & N & ... & ... \\
2012 Dec 7 & 09:02 & B4.8 & N18 W78 & 11621  & & Limb & ... & ... & & ... & ... & ... & ...  \\
2012 Dec 7 & 21:13 & C3.9 & N15 W73 & 11621  & & Limb & ... & ... & & ... & ... & ... & ...  \\
2013 Jan 4 & 08:22 & C1.3 & S17 W34 & 11639 & & 502 & NE    & N  & & 2 & N & A & ... \\
2013 Jan 5 & 16:19 & C1.7 & S13 W26 & 11645 & & 507 & SW    & N  & & 1 & N & A & ... \\
2013 Jan 6 & 06:20 & C8.4 & S16 W52 & N-AR  & & 475 & N     & N  & & 1 & N & A & ... \\
2013 Jan 9 & 14:37 & C3.1 & N30 W105 & 11640  & & Limb & ... & ... & & ... & ... & ... & ...  \\
2013 Jan 11 & 08:43 & M1.2 & N05 E36 & 11654 & & 768 & SW   & D  & & 1 & Y & B & ... \\
2013 Jan 12 & 06:10 & ... & S13 W115 & ...   & & Limb & ... & ... & & ... & ... & ... & ...  \\
2013 Jan 13 & 08:35 & M1.7 & N18 W22 & 11652 & & 747 & N    & N  & & 1 & Y & A & ... \\
2013 Jan 18 & 16:50 & C5.8 & N12 E30 & N-AR  & & 529 & W    & N  & & 1 & Y & B & ... \\

\enddata

\tablenotetext{a}{N: No clear front, D: Dark front, B: Bright front}
\tablenotetext{b}{(1) Asai et al. (2012); (2) Chen \& Wu (2011); (3) Chen et al. (2011);
(4) Cheng et al. (2012); (5) Dai et al. (2012); (6) Downs et al. (2012); 
(7) Gopalswamy et al. (2012); (8) Gopalswamy et al. (2013); (9) Gosain \& Foullon (2012);
(10) Harra et al. (2011); (11) Kozarev et al. (2011); (12) Kumar \& Manoharan (2013);
(13) Kumar et al. (2013); (14) Li et al. (2012); (15) Liu et al. (2010); 
(16) Liu et al. (2011); (17) Liu et al. (2012); (18) Long et al. (2011b); 
(19) Ma et al. (2011); (20) Olmedo et al. (2012); (21) Patsourakos et al. (2010);
(22) Schrijver et al. (2011);
(23) Shen \& Liu (2012a); (24) Shen \& Liu (2012b); (25) Shen \& Liu (2012c);
(26) Veronig et al. (2011); (27) Zheng et al. (2012a); (28) Zheng et al. (2012b)}

\end{deluxetable}

\clearpage
\hoffset=0.0in  

\onecolumn

\begin{deluxetable}{rrrrrrrrrrrr}
\tabletypesize{\scriptsize}

\tablecaption{Number of LCPFs included in the present study,
divided into different associations} 
\tablecolumns{12} \tablewidth{0pt}

\tablehead{
\colhead{} & \multicolumn{4}{c}{CME level} & \colhead{} & 
\multicolumn{2}{c}{Type II burst} & \colhead{} &
\multicolumn{3}{c}{171~\AA\ front} \\ 
\cline{2-5} \cline{7-8} \cline{10-12}\\
\colhead{Flare class} & \colhead{1} & \colhead{2} & \colhead{3} &
\colhead{4} & \colhead{} & \colhead{No} & \colhead{Yes} & 
\colhead{} & \colhead{Not clear} & \colhead{Dark} & \colhead{Bright}}

\startdata 
 X (11)    &  0 &  0 &  8 &  3 & & 1  & 10 & &  3 &  4 &  4 \\  
 M (57)    & 13 & 14 & 25 &  5 & & 19 & 38 & & 20 & 25 & 12 \\  
 C (49)    & 22 & 19 &  8 &  0 & & 28 & 21 & & 24 & 20 & 5 \\  
 $<$C (21) & 13 &  6 &  2 &  0 & & 15 & 6 & & 11 &  8 & 2 \\  
\enddata
\end{deluxetable}

\end{document}